# Stability and proton exchange in Te(OH)$_6$

Gabuda S.P., Kozlova S.G., Panov V.V., Golenkov E.O., Slepkov V.A.

Nikolaev Institute of Inorganic Chemistry, Russian Academy of Sciences, Acad. Lavrentiev Av. 3, 630090 Novosibirsk, Russia

Abstract. We consider the problem of mobility and localization of hydrogen atoms in the Te(OH)$_6$ crystals. The model of particles moving in a rapidly oscillating field was used.

Modern methods for studying vibronic coupling are mostly based on the approaches to the effect of Jahn-Teller electronic degeneration (effect and pseudo-effect). According to these approaches, displacement of heavy atoms and structural distortion in high symmetry systems are caused by the interaction between the ground and the nearest excited electronic states. The Möller-Plesset quantum-chemical method makes another approach for studying vibronic interactions to take into account higher-order electronic excitations in multielectronic systems [1]. Moreover the combine of the Möller-Plesset quantum-chemical method with the model moving of particle in a rapidly oscillating field [2] the mobility and localization of hydrogen atoms, phase transition, structural distortion in the crystals H$_2$C$_4$O$_4$ were investigated [3]. Here we consider the problem of mobility and localization of hydrogen atoms in the crystals of the acid Te(OH)$_6$.

H$_6$TeO$_6$ [or Te(OH)$_6$] crystals are characterized by polymorphism and have molecular structure type as for monoclinic modification (space group P2$_1$/$n$ with Z = 4 molecules in a cell), and for cubic modification (space group $F$4$_1$32; Z = 32). Molecule Te(OH)$_6$ has the deformed structure of an octahedron with O$_i$-Te-O$_j$ valent angles not equal 90$^o$. The feature of the deformed structure of molecules Te(OH)$_6$ remains not clear. Other feature is absence of hydrogen diffusion down to temperature of melting (395 K) [4,5].

On the other hand, similar molecules – molybdic (H$_6$MoO$_6$, H$_4$MoO$_5$ etc.), tungstic (H$_4$WO$_5$, H$_6$WO$_6$ etc.) are qualified as virtual objects. They exist only in the form of crystallohydrates (MoO$_3\cdot$nH$_2$O and WO$_3\cdot$nH$_2$O, n$\approx$1;2). This distinction associates with distinction of the d-orbitals filling: the d-orbitals of ions Te$^{6+}$ are filled [configuration Te (4d$^{10}$)], whereas 4d-and 5d-orbitals of ions Mo$^{6+}$ and W$^{6+}$ [configurations Mo(4d$^0$) and W(5d$^0$)] remain vacant.

*Computational and experimental details*

The electronic structure, the bonding energy, and the geometry of M(OH)$_6$ (M = Te, Mo, W) molecules were studied by the *ab initio* RHF (restricted Hartree-Fock) and LMP2 (second-order local Möller-Plesset) methods by the program Jaguar 5.5 [6]. The atomic basis 6-31G (p,d) was used for H and O [7], and the quasi-relativistic basis Lavcp [8] for the heavy atom Te, Mo and W. The bonding energy of the M(OH)$_6$ molecules formation was determined as $\Delta = E - \Sigma$, where E is the energy of the M(OH)$_6$ molecule and $\Sigma$ is the total energy of the atoms $M^0$, $6O^0$, and $6H^0$.

The mobility of hydrogen atoms was studied by the $^1$H NMR method in the Te(OH)$_6$ polycrystals. $^1$H NMR spectra were registered by a Bruker 300 NMR spectrometer in a temperature range of 160-400 K.

*Computing results*

Computing by Jaguar 5.5 package used the $O_h$ symmetry as the starting geometry for a M(OH)$_6$ molecules. In the Hartree-Fock approximation, the $O_h$ symmetry persisted during the geometry optimization. Computing by the LMP2 method showed that the $O_h$ symmetry of M(OH)$_6$ transformed into the $C_1$ symmetry, in compliance with experimental data of Te(OH)$_6$ structure (Table 1, Fig. 1). The contributions of electronic correlations to the bond energy ($\delta$) of the deformed M(OH)$_6$ molecule are represented in Table 2.

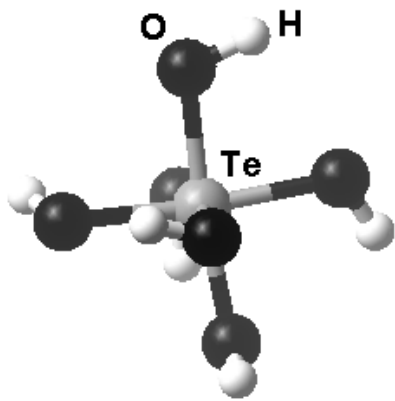

Figure 1. *The structure of Te(OH)$_6$ molecule computed by LMP2 method.*

Table 1. Geometrical parameters of the M(OH)$_6$ (M=Te,Mo,W) molecules (experiment and calculation)

| Bond | R, Å (calc.) | | | R, Å (exp.) |
|---|---|---|---|---|
| | Te(OH)$_6$ | Mo(OH)$_6$ | W(OH)$_6$ | Te(OH)$_6$ |
| <M – O$_i$> | <1.910> ± 0.006 | <1.955> ± 0.032 | <1.935> ± 0.026 | <1.909> ± 0.002 |
| <O-H$_i$> | <0.968> ± 0.001 | <0.973> ± 0.001 | <0.969> ± 0.001 | <0.985> ± 0.004 |
| Angle | Angle, gr. (calc.) | | | Angle, gr. (exp.) |
| | Te(OH)$_6$ | Mo(OH)$_6$ | W(OH)$_6$ | Te(OH)$_6$ |
| <O$_i$-M-O$_j$> | <172>± 2 | <159>±5 | <162>±4 | 176±1 |
| <O$_i$-M-O$_j$> | <90>±4 | <91>±10 | <91>±8 | <91>±3 |
| <H$_i$- O$_j$ – M> = φ$_o$ | <110>±1 | <111>±2 | <114>±2 | <115>±5 |

*A model of hydrogen atoms moving*

To investigate the effect of electronic correlations on the dynamics of hydrogen atoms, we considered a problem of particles moving in a rapidly oscillating field. The effective potential of energy for this system is given by [2]:

$$U_{eff} = U_1 + U_2 = A[-\cos(\varphi-\varphi_o)+\pi^2((d \cdot \nu_{ocs})/(D \cdot \nu_{libr}))^2 \sin^2(\varphi-\varphi_o)], \qquad (1)$$

where A is the energy constant which depends from environment; $\nu_{libr} = 4,2 \cdot 10^{13}$ Hz is the libration frequency of OH- groups, $\nu_{ocs}$ is the frequency of electronic correlations defined by correlation interactions ($\nu_{ocs} = \delta/n \sim 10^{15}$ Hz, where n = 6 - number of OH-groups, Table 2), d is the amplitude of fluctuations of electronic density for the oxygen atoms, D is the O-H distance, φ is the angle of deflection from the equilibrium, φ$_o$ is the equilibrium angle M-O-H (Table 1). Valent oscillations of OH-groups were neglected. U$_1$ in (1) may be considered as the analog of the Hartree-Fock potential, U$_2$ describes the influence of the electronic correlations on the mobility of hydrogen atom. Table 3 images specific parameters of the model. U$_{eff}$ shows one equilibrium position for the hydrogen atom of Te(OH)$_6$ molecule and two equilibrium positions for the hydrogen atoms of Mo(OH)$_6$ and W(OH)$_6$ molecules (Fig. 2). The possibility of hydrogen diffusion is testified by the double minima potential curve. Therefore the proton exchange is conceivable for Mo(OH)$_6$ and W(OH)$_6$ and the proton exchange is not conceivable for Te(OH)$_6$.

Table 2. The bond energies (eV) of the M(OH)$_6$ molecules, δ = LMP2-RHF

|  | LMP2 | RHF | δ | δ/n |
|---|---|---|---|---|
| Te(OH)$_6$ | -58.4 | -47.4 | -11.0 | -1.83 |
| Mo(OH)$_6$ | -69.5 | -51.0 | -18.5 | -3.08 |
| W(OH)$_6$ | -72.1 | -56.2 | -15.9 | -2.65 |

Table 3. The specific parameters U$_{eff}$ for M(OH)$_6$ molecules

| Molecule | $\nu_{ocs} \cdot 10^{15}$, Hz | d, Å | D, Å | 180-$\varphi_o$, gr |
|---|---|---|---|---|
| Te(OH)$_6$ | 0.44 | 0.006 | <0.968> | <70> |
| Mo(OH)$_6$ | 0.75 | 0.032 | <0.973> | <69> |
| W(OH)$_6$ | 0.64 | 0.026 | <0.969> | <66> |

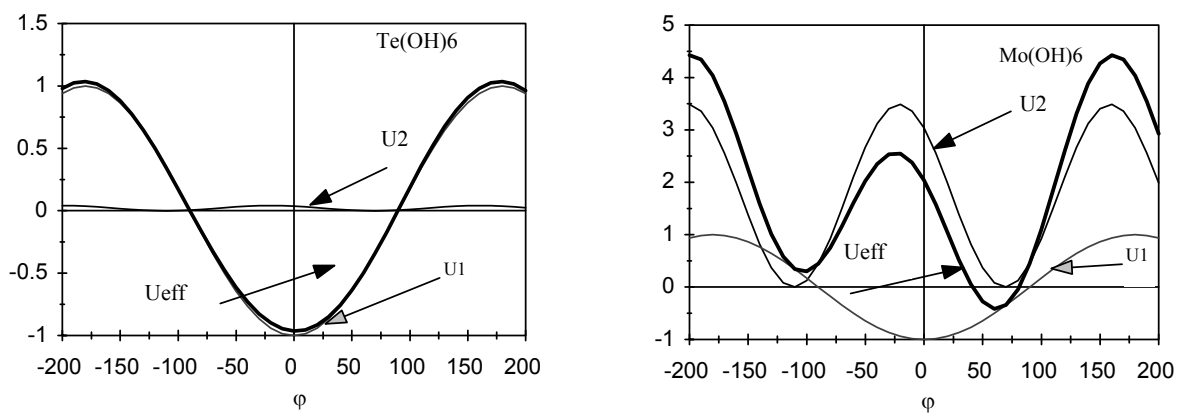

Figure 2. *Effective potential of energy U$_{eff}$ for fluctuations of OH-groups in the M(OH)$_6$ molecules. U$_1$ is the first item of U$_{eff}$ in the absence of electronic correlations; U2 is the additive due to electronic correlations.*

*$^1$H NMR data*

Figure 3 images $^1$H NMR data including experimentally registered second moments. The slight decrease in the $^1$H NMR second moments (or line width) with temperature is indicative the absent of the diffusion of hydrogen atoms. This is agreement with the calculated model.

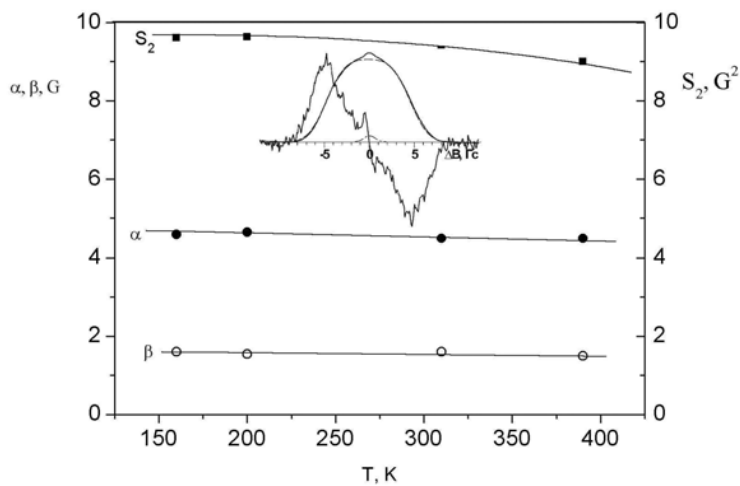

*Figure 4. Temperature dependences of the second moment ($S_2$) and Abraham function parameters ($\alpha$ and $\beta$ [9]). The inset images experimental $^1H$ NMR spectra (absorption and its derivative).*

The study was supported by grant 05-03-32263 from the Russian Foundation of Basic Research.